# Frequency conversion of abruptly autofocusing waves

Wei Gao,[1] Dong-Mei Wang,[1] Hai-Jun Wu,[1] Dong-Sheng Ding,[1,3] Carmelo Rosales-Guzmán,[1,2] and Zhi-Han Zhu,[1,*]

[1] *Wang Da-Heng Center, Heilongjiang Key Laboratory of Quantum Control, Harbin University of Science and Technology, Harbin 150080, China*

[2] *Centro de Investigaciones en Óptica, A.C., Loma del Bosque 115, Colonia Lomas del campestre, 37150 León, Gto., Mexico*

[3] *CAS Key Laboratory of Quantum Information, University of Science and Technology of China, Hefei, 230026, China*

Abruptly autofocusing waves and associated ring-Airy (RA) beams are attracting increasing interest owing to their fascinating properties such as their ability of abruptly autofocusing to small F-number. Optical frequency conversion via nonlinear interactions can further expand their applications to new area, yet are rarely studied. In this work, we report the frequency conversion of RA beams via sum-frequency generation using perfect flattop and common Gauss beams as the pump beams. The nonlinear transformation of the spatial complex amplitude of the signal and associated influences on autofocusing behavior, under different conditions of interaction location (i.e., original, autofocusing, and Fourier planes) and pump structure, were systematically studied and experimentally investigated. This proof-of principle demonstration provides a general guideline to build the frequency interface for abruptly autofocusing waves and a reference for relevant studies involving nonlinear transformation of abruptly autofocusing waves.

Accelerating waves and various Airy beams have attracted significant attention over the past years for breaking the paradigm of light propagating along a straight trajectory [1]. The term Airy beams originates from the fact that the beam's spatial amplitude is given in terms of an Airy function, which belongs to a shape-preserving solution of the potential-free Schrödinger equation, first revealed by Berry and Balazs in 1976 [2]. For paraxial beams, this means that they could resist diffraction upon propagation, and what is more attractive, have an ability to self-accelerate in the absence of any external potential. However, like other non-diffraction beams [3], the ideal Airy wave packets have infinite energy, and thus their experimental realization is physically impossible. To overcome this issue, according to the theory of Gori *et al*. on non-diffraction beams with finite apertures [4], so-called truncated Airy beams with finite energy were proposed and observed successfully in 2007 [5, 6]. These abbreviated but realizable Airy beams still inherits the main properties of the ideal ones, i.e., diffraction free and accelerating in the transverse direction over several diffraction distances [7, 8]. Furthermore, one can conveniently generate and control them via computer-generated holography, specifically, by imprinting a customized cubic phase on Gauss beams and taking its Fourier transformation. Truncated Airy beams have been widely studied through various subfields of optics, such as in the areas of microscopy, particle manipulation, laser machining, and plasmonic energy routing, among others [9–14].

Notably, the self-accelerating phenomenon and corresponding beams also exist in higher-dimension configurations, and the counterparts in cylindrical coordinates are usually referred to as cylindrically symmetric Airy beams, or ring-Airy (RA) beams. The outstanding propagation features of Airy wave packets—i.e., self-acceleration and non-diffraction—for the RA beams manifest in a unique autofocusing trajectory. Importantly, this autofocusing is fundamentally different from the ordinary Gauss beam focused by a lens (see Fig. 1(b) for an example). Specifically, in the propagation, the initial profile of an Airy ring can be maintained over several Rayleigh lengths, after which an abrupt autofocusing takes place at a focal plane, where the peak intensity can sharply increase by several orders of magnitude. This fascinating feature makes the RA beam a promising candidate for applications that strive to excite a certain light-matter interaction suddenly at a specific plane and avoid high radiation power elsewhere. Such applications include, for instance, laser ablation, laser surgery, optical tweezers, and laser-induced THz emission [15–17], among which the focusing properties, including the spot size, location, and intensity contrast are of great importance. All these properties, in both paraxial and nonparaxial domains, can be controlled fully by engineering initial beam parameters (e.g., phase and radial profiles, vortex orders, beam size, *etc*.) [18].

---

In experiments, The RA beam is usually generated by transforming the TEM$_{00}$ laser via computer-generated holography, where the key apparatus, e.g., the spatial light modulator (SLM), can only shape the spatial structure of the laser beam with a predefined wavelength. A common approach to reset the wavelength of structured light already generated is using optical frequency conversion, such as sum-/down-frequency generation (SFG/DFG) and four-wave mixing [19–23]. For RA beams, a key issue is to control the transformation of spatial complex amplitude during the nonlinear interaction and associated impacts on the autofocusing characteristics. Recently, it was shown that second-harmonic generation (SHG), i.e., a special self-pumped SFG, of an RA beam can inherit the autofocusing behavior of the fundamental wave [24]. Because the phase factor controlling the autofocusing behavior can be preserved in the square of the complex amplitude of RA beams, but only weighting quarter energy. However, it is worth noting that, first, SHG and high-order harmonic generation can only provide integer multiples of the input wavelength; second, the weight of the autofocusing component in the harmonic wave decreases exponentially with the order of harmonic waves. To date, the general frequency conversion for abruptly autofocusing waves via SFG/DFG, which enables a more flexible frequency modulation, is yet to be presented. In this work, we study SFG of abruptly autofocusing beams both theoretically and experimentally, where the RA beams (ultrafast pulses) in the near-infrared wavelength are upconverted to ultraviolet radiation. The nonlinear transformation of the spatial complex amplitude of signals and associated influences on their autofocusing behavior, under different conditions of interaction location, i.e., original ($z_0$), autofocusing ($z_f$), and Fourier ($z_\infty$) planes, and pump structure, were systematically studied.

The spatially complex amplitude of the original Airy ring at $z_0$ plane in cylindrical coordinates can be expressed as [14, 25]

$$u_{RA}(r,\varphi,z_0) = Ai\left(\frac{r_0-r}{w}\right) r^\ell \exp\left(\alpha\frac{r_0-r}{w} + i\ell\varphi + iv\frac{r_0-r}{w}\right), \quad (1)$$

where Ai(·) represents the Airy function, $\alpha$ is a truncation parameter, $r_0$ ($w$) is the radius (width) of the main (inner) ring, and $\ell$ is the azimuthal index of twisted phase corresponding to photonic OAM. In addition, $v$ is a parameter determining the initial launch angle of the parabolic trajectory, given by $\theta = v/(kx_0)$, where $k$ is the wavenumber, and $x_0$ is an arbitrary transverse scale. By assuming $\theta = 0$ and $\ell = 0$, the position of the autofocusing plane takes the form [26]

$$z_f = 4\pi\frac{w}{\lambda}\sqrt{R_0 w}. \quad (2)$$

Here, $R_0 \cong r_0 + w$ is the radius of the peak intensity of the main ring that is known exactly from the first zero of $u_{RA}'(r,\varphi,z_0)$. Not that the autofocusing position does not appear at the Fourier plane of the original Airy ring shown in Eq. (1).

The propagation trajectory of the RA beam from $z_0$ plane to the far field $z_\infty$ (equivalent to the focusing plane of a Fourier lens) can be numerically predicted by using the Collins diffraction integral [27, 28], which in cylindrical coordinates is given by

$$u_{RA}(r,\varphi,z) = \frac{i}{\lambda z}\exp(-ikz)\int r_0 dr_0 \int d\varphi_0 u_{RA}(r_0,\varphi_0,z_0) \\ \exp\left\{-\frac{ik}{2z}[r_0^2 - 2rr_0\cos(\varphi-\varphi_0) + r^2]\right\}. \quad (3)$$

Figure 1(a), for instance, shows the theoretical and observed propagation trajectory of a RA beam with parameters of $a = 0.07$, $r_0 = 0.28$, $w = 0.055$, and $R_0 = 0.332$. The Airy ring retains its radial profile over several Rayleigh distances and gradually contracts towards the axis, then, abruptly collapses on the axis at the autofocusing position ($z_f \approx 116.7$ mm) that coincides exactly with the prediction given by Eq. (2). Figure 1(b) shows the intensity contrast, defined by the ratio $I(z)/I(z_0)$ of the RA beam, where the green dashed curve is an equivalent Gauss beam focused by a lens with the same focal length and ratio (i.e., $I(z_f)/I(z_0)$). We see that, compared with equivalent Gauss beam, the autofocusing Airy ring has a significant advantage in focal abruptness, making it useful for applications requiring a short focal volume.

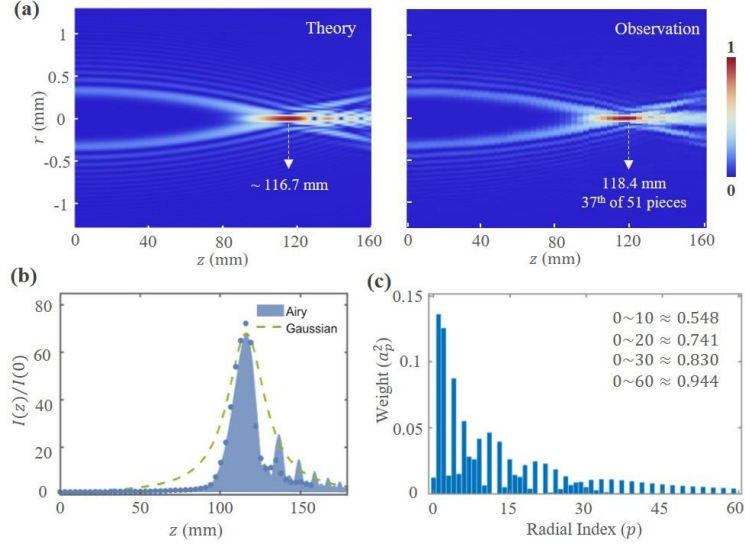

FIG. 1. (a) Propagation trajectory of a RA beam. (b) Intensity contrast versus z, where blue shadows (dots) represent theoretical (experimental) values, and the green dashed curve depicts an equivalent Gaussian beam for comparison. (c) Corresponding LG spectra, where upper right numbers are the weights of the top 10 ~ 60 components.

A crucial mechanism for the phenomenon of abrupt autofocusing is that the transverse profiles of RA beams, unlike Gaussian and other eigenmodes, are not maintained constant upon propagation. Therefore, the autofocusing phenomenon can also be interpreted from another perspective. The propagation variant RA beams can naturally be regarded as a superposition of eigenmodes of the paraxial wave equation [29, 30]. Indeed, Eq. (1) can be represented as a superposition of Laguerre-Gauss (LG) modes (see supplementary material for details), which are eigenmodes in the same coordinates, with a predefined azimuthal index ($\ell$) and various radial indices ($p$), given by

$$u_{RA}(r,\varphi,z_0) = \sum_p a_p^{z_0} \mathrm{LG}_p^\ell(r,\varphi,z_0), \ p \in \{0,1,2,...\} \quad (4)$$
$$a_p^{z_0} = \iint u(r,\varphi,z_0) LG_p^{-\ell}(r,\varphi,z_0) r dr d\varphi$$

where the weights $a_p^{z_0}$ denotes LG-mode spectra. Considering that diffraction is an adiabatic process, the modulus of each $a_p^{z_0}$ must be constant as the Airy ring proceeds to the far field. However, there is a phase factor that is added onto each of them, i.e., the Gouy phase given by $\phi_g = (2p+|\ell|+1)\arctan(z/z_R)$ for LG modes, and the propagating RA beam can therefore be expressed as

$$u_{RA}(r,\varphi,z) = \sum_p a_p(z) \mathrm{LG}_p^\ell(r,\varphi,z)_\perp, \ p \in \{0,1,2,...\} \quad (5)$$

where $a_p(z) = a_p^{z_0} \exp[i\phi_g(z)]$, and $\mathrm{LG}_p^\ell(r,\varphi,z)_\perp$ denotes the transverse complex amplitude (without Gouy phase) of the LG mode at the plane $z$. This equation indicates that, despite all the LG components enlarging at the same pace, the global pattern would still depend on the propagation due to the continuous change of their intramodal phases. According to this perspective, the mechanism of autofocusing can be considered as an intramodal interference of spatial modes. Hence, its underlying principle is similar with the Aharonov–Bohm effect and belongs to a geometric-phase induced phenomenon [31]. Figure 1(c) shows the LG spectra of the RA beam in Fig. 1(a), where the weights of the top 30 and 60 components are 83% and 94%, respectively.

To study the frequency conversion of abruptly autofocusing waves, one can start by exploring the nonlinear transformation of the spatial complex amplitude of RA signals and associated impacts on the resulting diffractive behavior. For SFG, the complex amplitude of the upconverted wave ($\omega_3$) in the interacting plane ($z_i$), denoted as $u_{SFG}^{\omega_3}(r,\varphi,z_i)$, is determined by the mixing wave of the pump ($\omega_2$) and the RA signal ($\omega_1$), i.e., $\kappa u_{RA}^{\omega_1}(r,\varphi,z_i) * u_p^{\omega_2}(r,\varphi,z_i)$, where $\kappa$ is the nonlinear coefficient of the medium, and phase matching requires a dispersion relation of $k(\omega_1) + k(\omega_2) = k(\omega_3)$. Because the transverse structure of the RA signal changes with $z$, we can infer that, for a given pump, the complex amplitude of generated SFG wave in the crystal would also vary. Here, we consider three cases in which the nonlinear interaction occurs at $z_0$, $z_f$, and $z_\infty$ planes of the RA signal, respectively, and analyze the corresponding impacts on the propagation trajectory of the upconverted RA beam. In addition, for simplicity, we assuming the interaction occurs in a short crystal and all used pumps have a plane wavefront in the crystal, such as far-field Gaussian beams, so that avoiding the influence of curvature radius of pumps on the signal. Thus, the SFG amplitude at $z_i$ plane for the three cases can be expressed as

$$\text{Case-1:} \quad u_{SFG}(r,\varphi,z_i)e^{-ik(\omega 3)z} \propto u_{RA}(r,\varphi,z_0) * u_p(r,\varphi)e^{-ik(\omega 1+\omega 2)z}, \quad (6)$$

$$\text{Case-2:} \quad u_{SFG}(r,\varphi,z_i)e^{-ik(\omega 3)z} \propto u_{RA}(r,\varphi,z_f) * u_p(r,\varphi)e^{-ik(\omega 1+\omega 2)z}, \quad (7)$$

$$\text{Case-3:} \quad u_{SFG}(r,\varphi,z_i)e^{-ik(\omega 3)z} \propto u_{RA}(r,\varphi,z_\infty) * u_p(r,\varphi)e^{-ik(\omega 1+\omega 2)z}, \quad (8)$$

and we next analyze their propagation trajectories through specific experiments.

In the experiment, for simplicity, we used the SFG of two ultrafast pulses at the same frequency, i.e., $\omega_1 = \omega_2$, which interact with each other non-colinearly in a type-I crystal. Figure 2(a) shows a schematic illustration of the setup, where the initial pulses (800 nm @ 120 fs and 80 MHz) were first divided into two paths, which were subsequently used as signal and pump beams, respectively. In both paths, we used a SLM to shape the spatial complex amplitudes of incident pulses [14, 32-35], see supplementary material for more details. Afterwards, the prepared signal and pump pulses were relayed by a 4f imaging system and then focused into a 0.5 mm $\beta$-BaB$_2$O$_4$ (BBO) crystal together non-colinearly. The generated SFG pulses were characterized by a CMOS-based beam profiler mounted on an optical rail with a step motor driver. We exploited three different imaging systems to observe the propagation trajectories of SFG waves from $z_0$ to the far field for cases in Eqs. (6)–(8), respectively, as shown in Fig. 2(b). Because the initial Airy ring of SFG waves, i.e., $u_{SFG}^{\omega 3}(r,\varphi,z_0)$, does not appear at the $z_i$ plane for the latter two cases. Specifically, the relation for case-3 is $u_{SFG}^{\omega 3}(r,\varphi,z_0) = \mathcal{F}\{u_{SFG}^{\omega 3}(r,\varphi,z_i)\}$; while in case-2 $u_{SFG}^{\omega 3}(r,\varphi,z_0)$ is a virtual image located at crystal's left side.

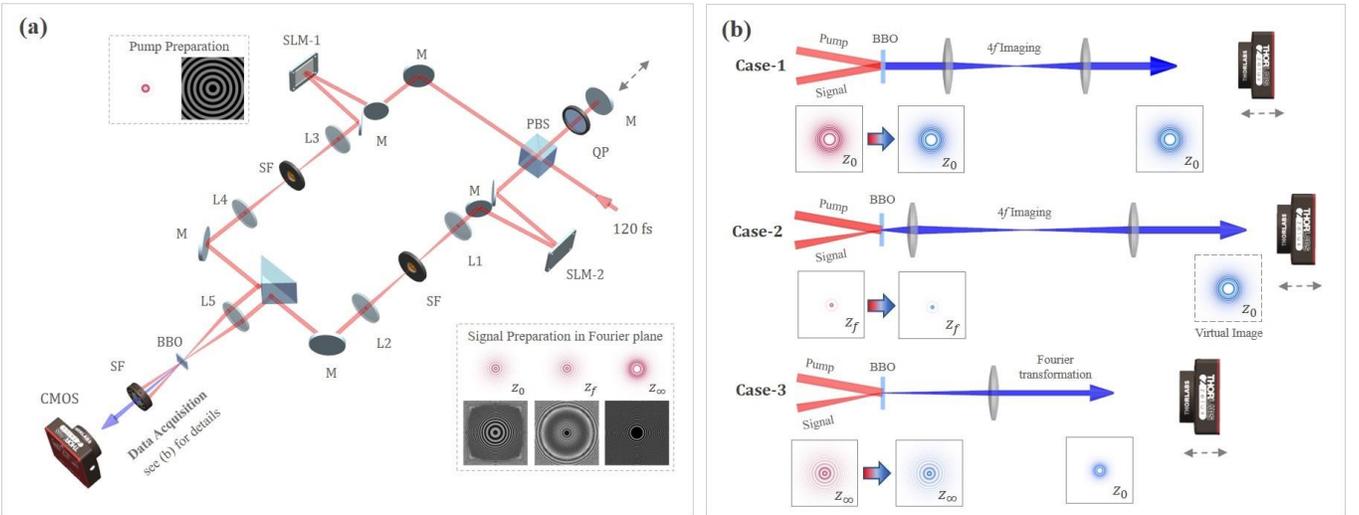

FIG. 2. (a) Diagram of the experimental setup, where the key components include the mirror (M), lens (L1-L5), polarizing beam splitter (PBS), quarter-wave plate (QP), spatial light modulator (SLM), spatial filter (SF), and beam profiler (CMOS). (b) Detailed imaging setup for propagation-trajectory tomography.

For convenience, the RA beam shown in Fig. 1 was chosen as the signal to be converted. We start from a relatively simple case — using a super-Gaussian mode, i.e., $u_{SG}(r,\varphi) = \exp[-(r/w)^n]$, with an order $n = 12$ and prepared via complex-amplitude modulation, as shown by the grating pattern near the SLM-2 in Fig. 2(a). Unlike the commonly used flattop-intensity beam obtained via phase-only modulation [36], although losing energy, the amplitude and phase of a super-Gaussian beam are both spatially uniform, whose complex amplitude can be efficiently cover RA signals at the focal region within the crystal (see Fig. S1 in supplementary material). In this approach, according to Eqs. (6)–(8), the spatial structure of signals is not affected by the pump during the frequency conversion. Figure 3(a) shows the simulated and observed propagation trajectories (or amplitude evolution) of upconverted RA pulses that correspond to the three cases shown in Eqs. (6)–(8), respectively. The observation confirms the theoretical prediction, where, for all cases, the transverse structure of RA signal at the interacting plane is efficiently maintained in the conversion. Thus, all the propagation profiles of upconverted waves were the same as that of the signal. This can be attributed to the fact that the super-Gaussian pump did not disturb the LG spectra shown in Fig. 1(c) during the interactions.

However, according to Eq. (2), the autofocusing position of the frequency converted signal would be changed due to the variation in wavelength. Here, owing to $\omega_3 = 2\omega_1$, the autofocusing positions of the SFG obtained in cases-1 and -2 should be double of the original signal. For case-3, in contrast, the autofocusing position of SFG would reduce to half of its original signal. This abnormal phenomenon is attributed to both the signal and the obtained SFG in case-3 being involved in a Fourier transformation implemented by using the lens. Specifically, as shown in Fig. 2(b), the beam profile of generated SFG in the crystal, i.e., $u_{SFG}^{\omega_3}(r,\varphi,z_i)$, completely inherits that of the signal at the Fourier plane, i.e., $u_{RA}^{\omega_1}(r,\varphi,z_\infty) = \mathcal{F}\{u_{RA}^{\omega_1}(r,\varphi,z_0)\}$, which is transformed by using a Fourier lens; thus, the initial Airy ring of the SFG, i.e., $u_{SFG}^{\omega_3}(r,\varphi,z_0)$, which was obtained using another Fourier lens with the same focal length, had a main-ring radius of $2r_0$; in consequence, according to Eq. (2), the final $z_f$ of SFG was reduced by half. Moreover, one can naturally reset the autofocusing length on demand by using a Fourier lens with a different focal length. All these predictions are also verified by the data in Fig. 3 (a). For more clearly, Fig. 3(b) exhibits maximum intensity as a function of $z$ for the original and upconverted signal, where blue shadows (dots) represent theoretical (experimental) values and green dashed curves are equivalently focused Gaussian beams for comparison.

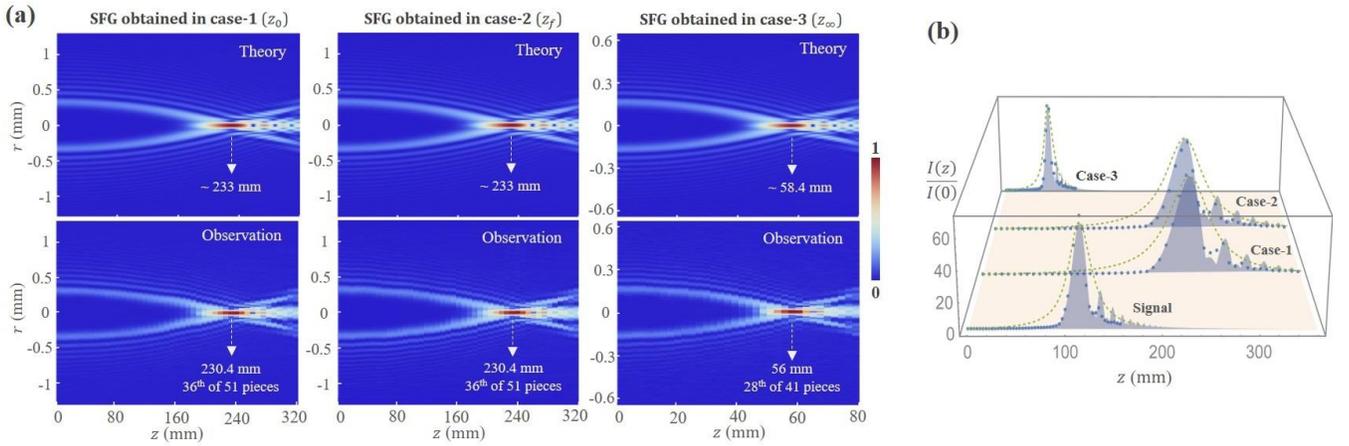

FIG. 3. (a) Propagation trajectories of upconverted beams via the SFG pumped by a perfect flattop pump, where the case-1, -2, and -3 correspond to the SFG occurring at $z_0$, $z_f$, and $z_\infty$ planes of the signal, respectively. (b) Intensity contrast of signal and three upconversions, where green dashed curves corresponding to equivalent Gaussian beams are shown for comparison.

Based on above results, we further consider the SFG pumped by a more commonly used Gaussian (TEM$_{00}$) beam. To avoid the spherical phase (i.e., curvature radius) introduced by the diffraction of pump, we used a far-field Gaussian beam as the pump in the crystal, i.e., $LG_0^0(r,\varphi,z_\infty)$. Under these circumstances, the beam size of the pump in the interaction plane, which played a pure spatial intensity modulator, controls the spatial spectra of obtained SFG waves. In the experiments, we chose 1-, 2-, and 3-times $R_0$ as the radius at half maximum of pump beams ($R_{HM}$), see Methods in supplementary material for more details. Figure 4(a) shows the maximum intensity as a function of $z$ for all cases and corresponding propagation trajectories are provided in supplementary material. In comparison with the SFG pumped by the super-Gaussian mode, (i) the autofocusing positions in all cases are roughly unchanged; (ii) as $R_{HM}$ increases, in all cases the autofocusing properties of upconverted beams tend to the results obtained using super-Gaussian pump, i.e., maintaining the focusing capability of the signal; and (iii) the focusing capability of upconverted beams, including focusing ratio and abruptness, decreases in both cases-1 and -3, yet seems to be enhanced in case-2.

Among these phenomena, phenomenon (i) is attributed to the far-field Gaussian pump with a plane wavefront that does not disturb the intramodal-phase structure and curvature radius of LG spectra of the signal in the frequency conversion. The reason behind phenomenon (ii) is the Gaussian-shaped amplitude modulation applied by the pump. This nonlinear amplitude modulation causes a low-pass filtering for the spatial spectra of upconverted waves and the degree is varied with $z$ [37, 38], (see the LG spectra shown in the data-2 in supplementary material). Regarding phenomenon (iii), the loss of high-order components inevitably leads to the variation in focusing property. For cases-1 and -3, the loss of high-order $LG_p^0$ components decreased the extinction ratio (or interference visibility) of the outer rings at the autofocusing plane, leading to the degradation of the focusing ratio, i.e., $I(z_f)/I(z_0)$. Case-2 is of particular interest, as we found that the focusing ratio seems to be enhanced with decreasing $R_{HM}$. This is because the Gaussian-shaped intensity modulation imposed by the pump cuts off the amplitude surrounding the focal spot. However, it is important to note that the focusing capability of the up-converted beam in this case is also degraded. This is because the upconverted beam here loses the key merit of RA beams. Figure 4(b) clearly shows the comparison of SFG and equivalent Gaussian and RA beams, which have the same focusing ratio and length. The comparison clearly tells us what is the price of enhancing focusing ratio costs, i.e., at the expense of focusing abruptness.

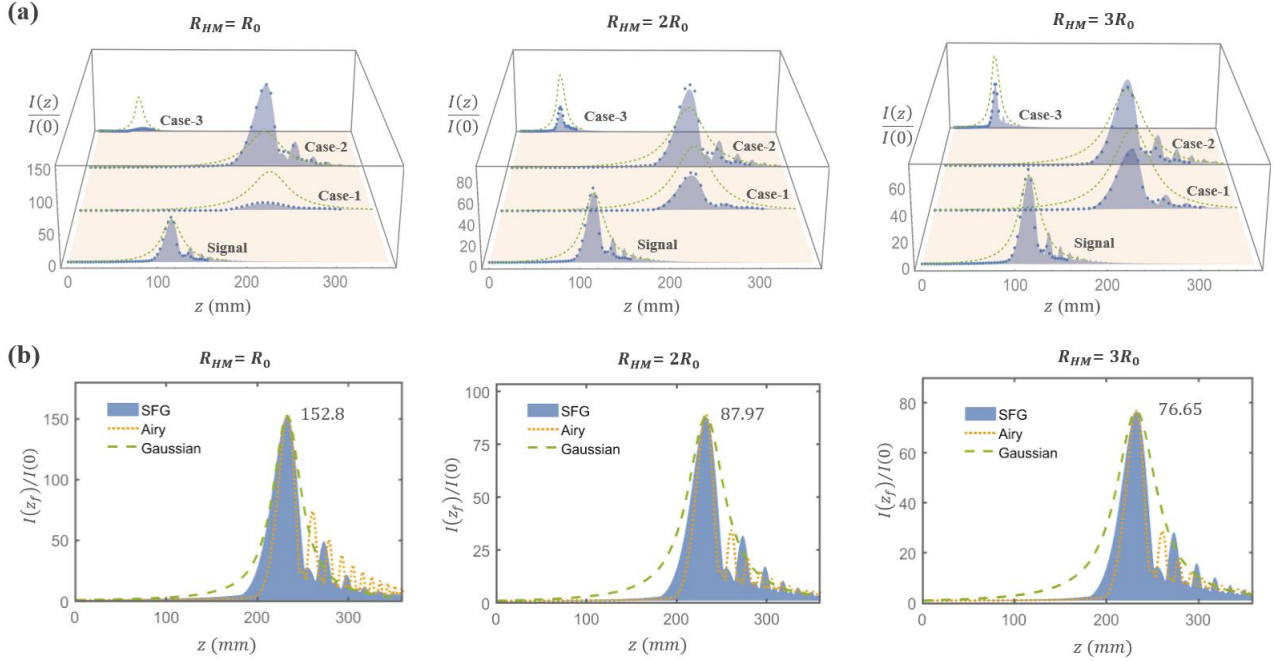

FIG. 4. (a) Intensity contrast of SFG pumped by Gaussian beams. (b) Intensity contrast of SFG obtained in case-2 and pumped by Gaussian beams with $R_{HM}$ = 1, 2, and 3 $R_0$, respectively, where orange and blue dashed lines are those of corresponding equivalent RA and Gauss beams for comparison.

In the above demonstration, we found that by using super-Gaussian beams as the pump, a high-fidelity frequency conversion is realized for RA beams, which inherits all the key focusing properties of input signals. Otherwise, for convenience, using a Gaussian beam with an appropriate waist likewise yields an acceptable result. Regarding the three interacting planes, the RA signal at the autofocusing plane has a smaller beam size, where one can use a small pump size to realize a high-fidelity and -efficiency conversion. If the application requirement is not frequency conversion, but shaping the region of nonlinear light-matter interactions, such as nonlinear optical microscopy, our results, particularly those obtained in the case-2, still maintain referential value. Although we used a special SFG in the experiment, i.e., $\omega_1 = \omega_2$, the verified theoretical methods obviously compatible with frequency-nondegenerate SFG and, according to Eq. (2), the main change would reflect in autofocusing positions. Besides, if a long interaction region is considered, the influence of Gouy phase of mixing waves on the interaction is worthy to further explore [23, 39,40]. In summary, we studied the frequency conversion of autofocusing RA beams both theoretically and experimentally. The nonlinear transformation of an RA beam in the SFG under different conditions of interaction location and pump structure, and the resulting diffractive behaviors were studied in detail. Our results provide a general guideline to build the frequency converter for RA beams, and contribute to other relevant studies involving nonlinear optics of abruptly autofocusing waves.

## SUPPLEMENTARY MATERIAL

See supplementary material for Methods (theoretical and experimental details) and Additional Data.

## ACKNOWLEDGMENT

We acknowledge funding from the National Natural Science Foundation of China (Grant Nos. 62075050, 11934013, 61975047, and U20A20218), the High-Level Talents Project of Heilongjiang Province (Grant No. 2020GSP12), and the National Key R&D Program of China (Grant No. 2017YFA0304800).

# Supplementary Materials for
# Frequency conversion of abruptly autofocusing waves

Wei Gao,[1] Dong-Mei Wang,[1] Hai-Jun Wu,[1] Dong-Sheng Ding,[1,3] Carmelo Rosales-Guzmán,[1,2] and Zhi-Han Zhu[1,*]

[1] *Wang Da-Heng Center, Heilongjiang Key Laboratory of Quantum Control, Harbin University of Science and Technology, Harbin 150080, China*
[2] *Centro de Investigaciones en Óptica, A.C., Loma del Bosque 115, Colonia Lomas del campestre, 37150 León, Gto., Mexico*
[3] *CAS Key Laboratory of Quantum Information, University of Science and Technology of China, Hefei, 230026, China*

## Methods

**Theoretical background.** The spatial complex amplitude of the LG mode used mentioned in the main text is given in cylindrical coordinates $\{r,\varphi,z\}$ by

$$LG_p^\ell(r,\varphi,z) = \sqrt{\frac{2p!}{\pi(p+|\ell|)!}} \frac{1}{w(z)} \left(\frac{\sqrt{2}r}{w(z)}\right)^{|\ell|} \exp\left(\frac{-r^2}{w_z^2}\right) \times L_p^{|\ell|}\left(\frac{2r^2}{w_z^2}\right) \exp\left[-i\left(kz + \frac{kr^2}{2R_z} + \ell\varphi - i\phi_g\right)\right], \quad \text{(S1)}$$

where $w_z = w_0\sqrt{1+(z/z_R)^2}$, $R_z = z^2 + z_R^2/z$, and $\phi_g = (2p+|\ell|+1)\arctan(z/z_R)$ denote the beam waist, radius of curvature, and Gouy phase upon propagation (here $z_R = kw_0^2/2$ is the Rayleigh length), respectively, and $L_p^{|\ell|}(\cdot)$ is the Laguerre polynomial with mode orders $p$ and $|\ell|$, given by $L_p^{|\ell|}(\gamma) = \sum_{k=0}^{p}\left[(|\ell|+p)!(-\gamma)^k\right] / \left[(|\ell|+k)!k!(p-k)!\right]$.

The transverse overlap between pumps (super-Gauss and common Gauss modes) and signals (the three cases) within the crystal were designed in below.

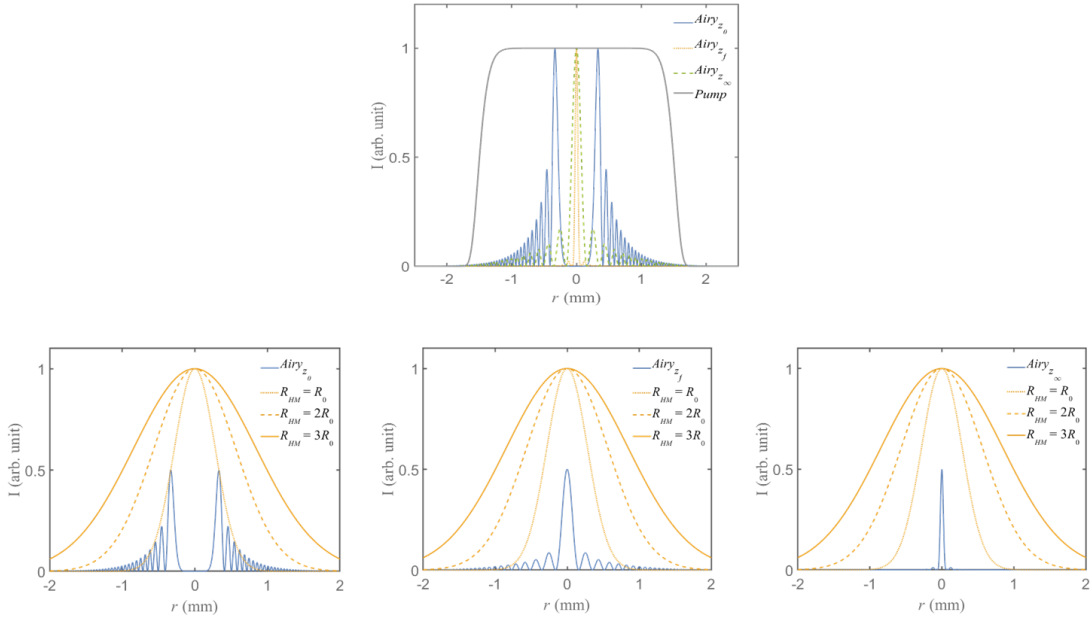

FIG. S1. Spatial overlap of pumps and signals at the interaction plane within the crystal designed for the experiment.

---

[*] zhuzhihan@hrbust.edu.cn

**RA beams generation via SLM.** The most common approach to generate the RA beam in experiments is to exploit SLM loading a hologram that corresponds to the complex amplitude of Eq. (1) or its Fourier transformation (a Fourier transformation is required). In experiments, the RA signals generated at the surface of SLM-2 (Fourier plane of the BBO crystal) for the case-1, -2, and -3 are shown in Fig. S2(a), (b), and (c), respectively. The curves in the second row are intensity mask, where OI and TI denote the transverse intensities of original $TEM_{00}$ light and target signals, respectively, and the third row are corresponding gratings.

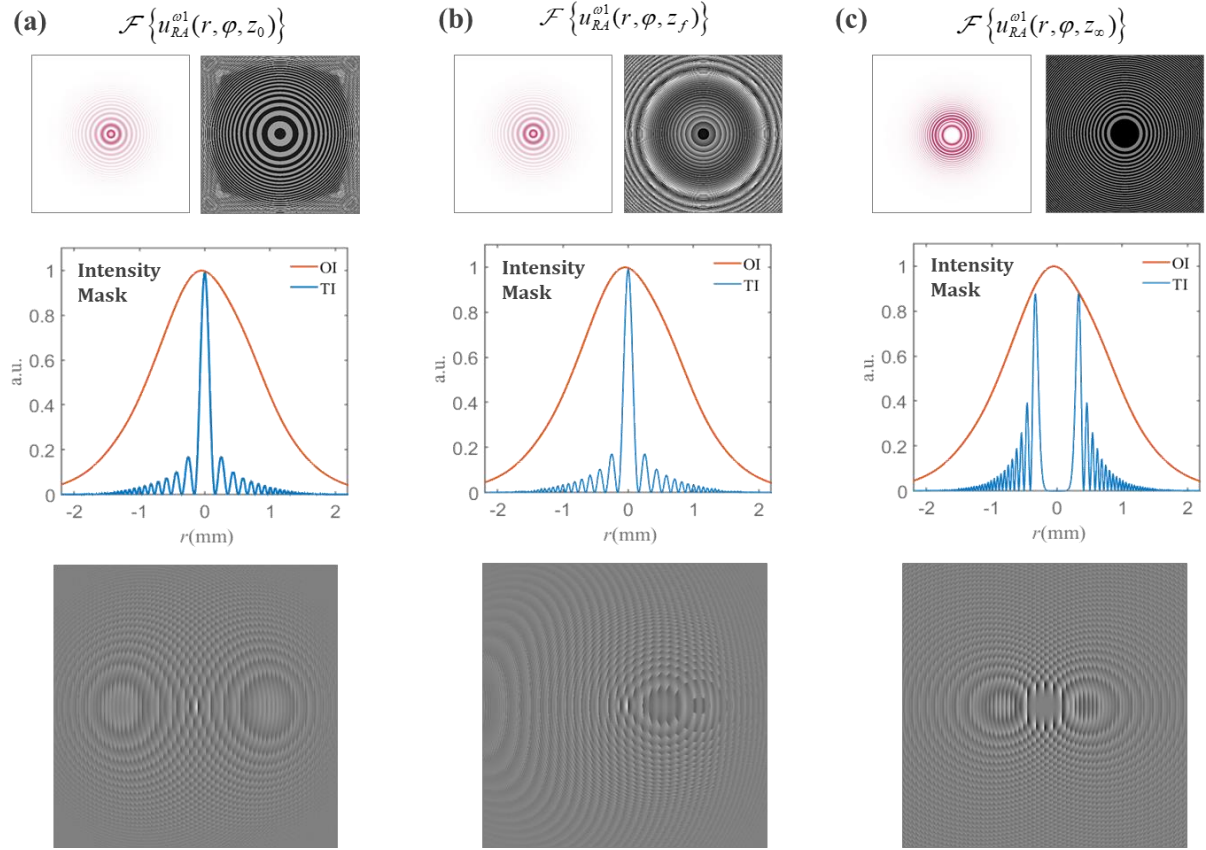

FIG. S2. Complex amplitude modulation for RA signals preparation.

## Additional Data

**Data-1.** Propagation trajectories (both theoretically predicted and experimentally observed) of upconverted beams via the SFG pumped by a Gaussian pump, where the case-1, -2, and -3 correspond to the SFG occurring at the original, autofocusing, and Fourier planes of the signal, respectively.

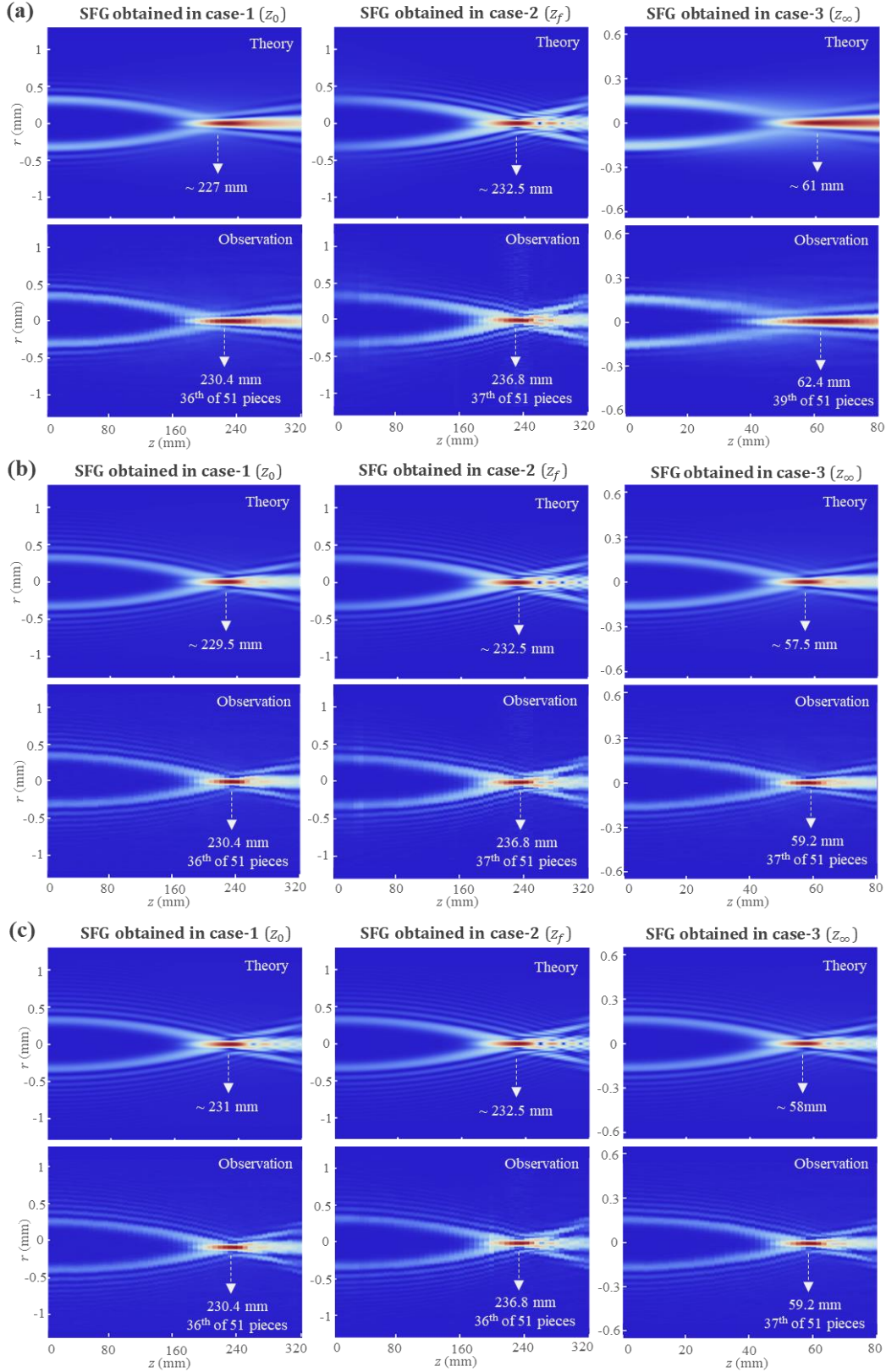

FIG. S3. Original data base (i.e., propagation trajectories) for the data shown in Fig. 4 (a) in the main text, where (a)-(c) correspond to $R_{HM}$ = 1, 2, and 3 $R_0$, respectively.

**Data-2.** Fig. 2S shows the LG spectra the SFG obtained in case-1, -2, and -3 with different pump beam sizes. We see that the weight of low-order components decreases with the $R_{HM}$ of the pump and gradually closes into that of the original signal (see Fig. 1 in the main text), i.e., gradually becomes the flattop pump. In addition, for a given $R_{HM}$, the degree of the filtering among the three cases is ranked by cases -2, -3, and -1, which is determined by the beam size of the signal at the interaction plane.

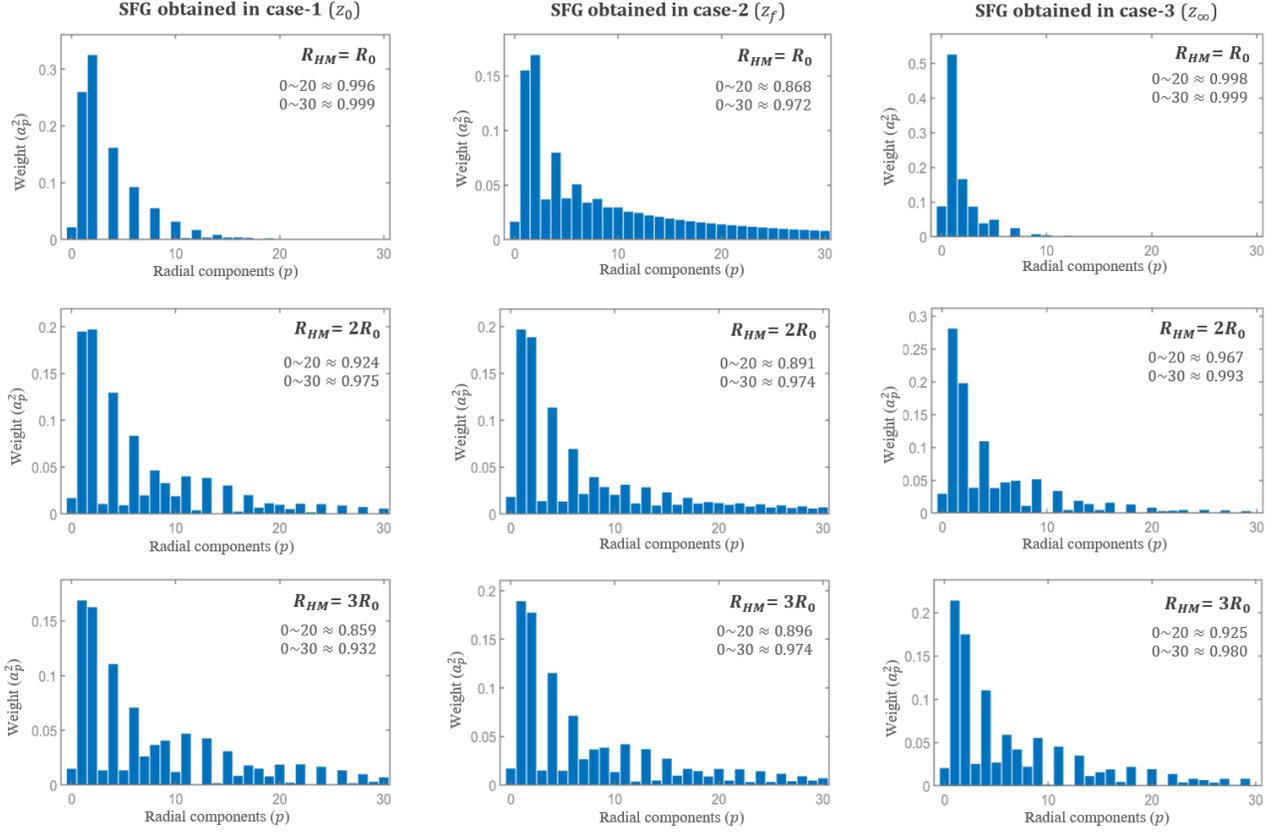

FIG. S4. LG spectra of the SFG obtained in case-1, -2, and -3 with different pump beam sizes.